\documentclass[twocolumn]{aastex63}
\usepackage{amsmath}
\usepackage{bm}

\received{today}
\submitjournal{ApJ}

\shorttitle{Fermi and eROSITA Bubbles as Persistent Structures of The MW Galaxy}
\shortauthors{Shimoda \& Asano}

\begin{document}

\title{Fermi and eROSITA Bubbles as Persistent Structures of the Milky Way}

\correspondingauthor{Jiro Shimoda}
\email{jshimoda@icrr.u-tokyo.ac.jp}

\author[0000-0003-3383-2279]{Jiro Shimoda}
\affiliation{Institute for Cosmic Ray Research, The University of Tokyo, 
5-1-5 Kashiwanoha, Kashiwa, Chiba 277-8582, Japan}

\author[0000-0003-4366-6518]{Katsuaki Asano}
\affiliation{Institute for Cosmic Ray Research, The University of Tokyo, 
5-1-5 Kashiwanoha, Kashiwa, Chiba 277-8582, Japan}




\begin{abstract}
The Fermi and eROSITA bubbles, large diffuse structures in our Galaxy,
can be the by-products of the steady star formation activity.
To simultaneously explain the star formation history of the Milky Way
and the metallicity of $\sim Z_\odot$ at the Galactic disk, a steady
Galactic wind driven by cosmic-rays is required.
For tenuous gases with a density of $\lesssim 10^{-3}$~cm$^{-3}$,
the cosmic-ray heating dominates over radiative cooling, and the gas
can maintain the virial temperature of $\sim0.3$~keV ideal for escape
from the Galactic system as the wind. A part of the wind falls back onto
the disk like a galactic fountain flow. We model the wind dynamics according
to the Galactic evolution scenario and find that the scale height and
surface brightness of the X-ray and the hadronic gamma-ray emissions from
such fountain flow region can be consistent with the observed properties of
the Fermi and eROSITA bubbles. This implies that the bubbles are
persistent structures of the Milky Way existing over (at least)
the last $\sim1$~Gyr, rather than evanescent
structures formed by non-trivial, $\sim10$~Myr past Galactic Center
transient activities.
\end{abstract}

\keywords{gamma-rays:ISM --- X-rays:ISM --- (ISM:) cosmic rays --- Galaxy: halo --- Galaxy: evolution --- ISM: jets and outflows}


\section{Introduction}
\label{sec:intro}
The Fermi and eROSITA bubbles (FBs and eRBs) are the largest gamma-ray and
X-ray emitting objects in the sky, respectively \citep{su10,su12,ackermann14,predehl20}.
They look like nearly symmetrical pairs of bubbles rising above and below the center of
our Galaxy. The FBs extend about $50^\circ$, and their emission mechanism is under debate, whether
the leptonic scenario \citep[inverse Compton scattering by relativistic electrons;][]{2011PhRvL.107i1101M,2019A&A...622A.203M,
2011ApJ...731L..17C,
2014ApJ...790...23C,
2015ApJ...804..135C,
sasaki15,2017ApJ...850....2Y}
or the hadronic
scenario \citep[decay of $\pi^0$ particles produced by collisions
between relativistic protons and target nuclei in thermal gas;][]
{2011PhRvL.106j1102C,
2015ApJ...808..107C,
2013ApJ...775L..20F,
2015ApJ...799..112C}. 
The eRBs extend up to $\sim80^\circ$ and are dominated by
thermal X-ray emission. In the last decade, most of the literature
suggested that the FBs and eRBs were formed by $1\mathchar`-10$~Myr
past Galactic Center (GC) burst-like
activities
\citep[e.g.,][, for recent studies]{blandh19,yang22,nguyen22,owen22,sarkar23,sarkar24},
implying that the bubbles are evanescent structures of the Milky Way Galaxy (MW).
However, the true formation mechanisms are still unknown.
\par
Linearly polarized radio observations have also reported such bilobal
giant structures \citep[][hereafter Giant Radio Lobes, GRLs]{carretti13}.
Several bright filamentary substructures in the GRLs trace the corresponding
parts of the FBs and eRBs very well. The polarization observations also show
that the magnetic fields above and below the Galactic disk are perpendicular
to the disk everywhere. These facts strongly support the existence of outflows
from the disk, however, the filamentary substructures may rule out a simple
bubble-like morphology of the outflow.
\par
The outflow from the Galactic disk is also strongly motivated to explain
the total amount of metals in the current disk, $\sim10^7~M_\odot$, which is
estimated with the typical metallicity of $Z_\odot\sim0.01$ and the total
mass of gaseous matter within the star-forming region of the disk $\sim10^9~M_\odot$
\citep[e.g.,][]{misiriotis06}. Without the outflow from the Galactic disk, the
metal amount would be much larger than the above estimate; The star formation
in the MW has continued at a rate of $\dot{M}_{\rm sf}\gtrsim3~M_\odot$~yr$^{-1}$
during the cosmic-age of $t_{\rm age}\sim14$~Gyr \citep{haywood16}. The Salpeter
initial mass function (IMF) gives a fraction of massive stars as
$f_{\rm ms}\sim0.1$ \citep[e.g.,][]{kroupa01,chabrier03}. From the ratio of the
metal mass of the supernova ejecta to the mass of the progenitor star as
$f_{\rm ej}\sim0.1$ \citep[e.g.,][]{sukhbold18,chieffi20}, we obtain the total
mass of metals ejected by the supernovae during the cosmic-age as
$M_Z\sim f_{\rm ej}f_{\rm ms}
\dot{M}_{\rm sf}t_{\rm age}\gtrsim4\times10^8~M_\odot\gg10^7~M_\odot$.
Thus, $\gtrsim97$\% of the ejected metals
should be removed from the disk by the outflow \citep{shimoda24}.
The observed mid-infrared large structure around the GC \citep{blandh03}
may be the evidence of the outflow that removes `missing' metals.
\par
The diffuse X-ray emission from hot gaseous matter with a temperature of
$\sim0.3$~keV including the eRBs \citep[e.g.,][]{kataoka18} may ensure
the existence of the outflow from the disk, however, the origin of
such diffuse X-ray emission is a long-standing problem \citep[e.g.,][reviews
for the emission from the disk]{koyama18}.
The possible existence of the Galactic wind driven by cosmic rays (CRs) 
was pointed out by \citet{breitschwerdt91} for the first time.
\citet{breitschwerdt99} included the effects of the radiative cooling into the wind model
and showed that the observed diffuse X-ray spectrum can be explained by the wind scenario.
\citet{everett08,evertt10}
also showed that the surface brightness profile can be fitted by the successful wind model.
\par
However, considering the cooling and CR effects with the realistic boundary condition, the
geometrical structure of the Galactic wind may be complicated. The idealized solution of
successful outflow could not apply to the entire region above the Galactic disks.
From observed OVI absorption lines, \citet{shapiro76} suggested the galactic fountain flow, in which ejected hot
gases above the disk cool radiatively and fall onto the disk. They estimated
the scale height of the fountain flow to be $\sim1$~kpc, which is too small to
explain the eRBs. \citet{shimoda22a} pointed out that the heating rate due to
cosmic rays (CRs) can be comparable to the radiative cooling rate for tenuous
gas, and the gas can escape from the Galaxy as the Galactic wind, whose number density
and temperature are $\sim10^3~{\rm cm^{-3}}$ and $\sim0.3$~keV, respectively.
\par
The wind model should be consistent with the star formation activity.
With the CR-driven wind scenario of \citet{shimoda22a}, in which both the cooling and CR diffusion effects are included, \citet{shimoda24} reproduced
the long-term evolution of the MW over cosmic time consistently with
the star formation rate (SFR) and metallicity, taking into account
the CR heating and stellar dynamics. They found that
the MW history can be self-consistently explained
if $\sim5$-$10$\% of the supernova explosion energy is used to drive the wind.
\citet{gupta23} reported that the metal abundance of the eRBs can be
consistent with the metal enrichment scenario by supernovae/stellar winds.
\par
As discussed in \citet{armillotta24} for example, dense cold clouds falling onto
the disk and hot gases launched from the disk coexist in the Galactic halo. Motivated
by the complicated structure of the wind and fountain flow with CRs, we study the X-ray
and hadronic gamma-ray emissions from the Galactic halo region without a short and intense
activity of the Galactic center in this paper. Being difficult to establish a fully
self-consistent model, as a first step, we adopt simplified schemes to calculate the
Galactic wind motion and CR distribution to roughly reproduce the Fermi and eROSITA bubbles.
\par
This paper is organized as follows: In Section~\ref{sec:model}, we describe
our model based on the long-term evolution model and observations of the MW.
The numerical results of our model are presented in Section~\ref{sec:results}.
The thermal X-ray and hadronic gamma-ray emissions from the Galactic halo are
reproduced. The implications of our scenario for future studies are discussed
in Section~\ref{sec:discussion}.

\section{Model description}
\label{sec:model}
To produce the sky intensity maps of the hadronic gamma-ray and thermal
X-ray photons, we consider the recent $\sim1$~Gyr evolution of the MW
according to the latest Galactic evolution model by \citet{shimoda24}.
The Galactic system is modeled under the approximation of axial symmetry
with the cylindrical coordinate of $(R,z)$. The SFR at the disk is
estimated from the disk gas distribution based on the HI and H$_2$ observations \citep{misiriotis06}. The CR energy density is estimated from the supernova
rate with the Salpeter IMF \citep[e.g.,][]{kroupa01,chabrier03}. The outflow
dynamics is regulated by the Galactic gravitational potential due to stars
\citep{miyamoto75} and the NFW-like dark matter (DM) distribution \citep{nfw96}.
The details of our method are as follows.
\par
\subsection{the Galactic disk}
\label{sec:disk}
\begin{figure}[htbp]
\includegraphics[scale=1.0]{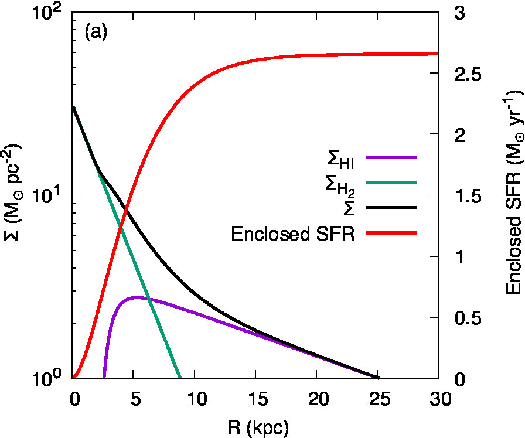}
\caption{
The surface mass density profiles of $\Sigma_{\rm HI}(R)$,
$\Sigma_{\rm H_2}(R)$, and $\Sigma(R)\equiv\Sigma_{\rm HI}
+\Sigma_{\rm H_2}$ (the left-hand side vertical axis).
The enclosed SFR, $2\pi\int_0^R \dot{\Sigma}_{\rm sf}(R')R'dR'$,
is also shown (the right-hand side vertical axis).}
\label{fig:sigma}
\end{figure}

We model the disk gas surface mass density following \citet{misiriotis06}.
The HI gas surface density is
%
\begin{eqnarray}
\Sigma_{\rm HI}=\Sigma_{0,{\rm HI}}
\exp\left[ -\frac{R}{R_{\rm HI}}-\left(\frac{R_{\rm t}}{R}\right)^4 \right],
\label{eq:HI]}
\end{eqnarray}
%
where $R_{\rm HI}=18.24$~kpc, $R_{\rm t}=2.75$~kpc, and the normalization
factor of $\Sigma_{0,{\rm HI}}$ is adjusted to satisfy that the enclosed
mass within $R=30$~kpc is $3.9\times10^9~M_\odot$ (the mass within $R=\infty$
is $8.2\times10^9~M_\odot$). Here we modify the cut-off shape as $\exp
\left[-(R_{\rm t}/R)^4\right]$ from a discontinuous cut-off originally assumed
to keep its differential finite. The H$_2$ gas surface density is given by
%
\begin{eqnarray}
\Sigma_{\rm H_2}(R)
=\Sigma_{0,{\rm H_2}}
\exp\left[ -\frac{R}{R_{\rm H_2}} \right],
\label{eq:H2}
\end{eqnarray}
%
where $R_{\rm H_2}=2.57$~kpc and the enclosed mass within $R=30$~kpc
is $1.3\times10^9~M_\odot$. As stars are formed in molecular
clouds, the surface SFR density can be written as
%
\begin{eqnarray}
\dot{\Sigma}_{\rm sf}(R)=\frac{\Sigma_{\rm H_2}}{\tau_{\rm sf}},
\label{eq:sfr}
\end{eqnarray}
%
where the effective gas consumption timescale due to the star formation
is assumed as $\tau_{\rm sf}=0.5$~Gyr \citep[see,][for details]{inutsuka15}.
Figure~\ref{fig:sigma} shows the surface
densities of $\Sigma_{\rm HI}$, $\Sigma_{\rm H_2}$, $\Sigma\equiv
\Sigma_{\rm HI}+\Sigma_{\rm H_2}$, and the enclosed SFR, $2\pi\int_0^R
\dot{\Sigma}_{\rm sf}(R')R'dR'$. The total SFR within $R=30$~kpc becomes
$2.76~M_\odot~{\rm yr^{-1}}$, which is consistent with the SFR in the MW
averaged for recent $\sim1$~Gyr \citep{haywood16}.
%
%
\par
In our model, the supernova explosions blow the gas off the disk.
The event rate density of supernovae is 
%
\begin{eqnarray}
\dot{N}_{\rm sn}(R)
=f_{\rm ms}\frac{ \dot{\Sigma}_{\rm sf}(R) }{ \bar{m}_{\rm *,ms} },
\label{eq:sn rate}
\end{eqnarray}
%
where the Salpeter IMF provides the average mass of supernova progenitors
$\bar{m}_{*,{\rm ms}}=30.9~M_\odot$ and the fraction of supernova progenitor
stars $f_{\rm ms}=0.22$ \citep[see,][for details]{shimoda24}. A fraction of
the supernova explosion energy is converted to the outflow energy so that
the surface density of the outflow rate $\dot{\Sigma}_{\rm blown}$ satisfies
%
\begin{eqnarray}
\frac{kT_{\rm w}}{m_{\rm p}}\dot{\Sigma}_{\rm blown}(R)
=\eta_{\rm blown}E_{\rm sn} \dot{N}_{\rm sn}(R),
\label{eq:kTw}
\end{eqnarray}
%
where $\eta_{\rm blown}=0.05$ is the outflow conversion efficiency, the
supernova explosion energy $E_{\rm sn}=10^{51}$~erg, the assumed temperature
of the outflow gas $T_{\rm w}=3\times10^6$~K \citep[e.g.,][]{kataoka13,nakashima18,shimoda22a}, $m_{\rm p}$ is the proton mass, and
$k_{\rm B}$ is the Boltzmann constant. As a result, the outflow rate from
the disk can be written as $\dot{\Sigma}_{\rm blown}=\eta_{\rm w}f_{\rm ms}
\dot{\Sigma}_{\rm sf}$, where $\eta_{\rm w}\equiv\eta_{\rm blown}m_{\rm p}
E_{\rm sn}/(\bar{m}_{*,{\rm ms}}k_{\rm B}T_{\rm w})\simeq3.3$.
\par
We assume that the blown gas extends with a scale height of
$z_{\rm hl}=2$~kpc, which corresponds to the height estimated in the classical
galactic fountain scenario \citep{shapiro76}. In our model, the fountain
regions are laid on the disk with the thickness of $z_{\rm hl}$ and treated
separately from the disk region with the thickness of $H=0.3$~kpc. We refer
to the fountain region as the layer in the following. The gas in the layer is
assumed to escape as the wind from a height of $z=z_{\rm hl}$ with a speed of
$C_{{\rm s,w}}=\sqrt{k_{\rm B}T_{\rm w}/(\mu m_{\rm p})}\simeq200~{\rm km~s^{-1}}
(T_{\rm w}/3\times10^6~{\rm K})^{1/2}$ to the halo, where $\mu=0.6$ is the mean
molecular weight. Then, the density of the wind is given under the steady-state
approximation as
%
\begin{eqnarray}
\rho_{\rm w}(R,z_{\rm hl})
&=&
\frac{ \eta_{\rm w}f_{\rm ms} }{2C_{\rm s,w}}
\dot{\Sigma}_{\rm sf}(R),
\label{eq:rho w}
\end{eqnarray}
%
where the factor of $2$ in the denominator represents the two layers
existing above and below the disk. The typical number density of the wind becomes
%
\begin{eqnarray}
n_{\rm w}\equiv\rho_{\rm w}/m_{\rm p}
\simeq0.4\times10^{-3}~{\rm cm^{-3}}
\left(\frac{ \Sigma_{\rm H_2} }{ 10~M_\odot~{\rm pc^{-2}} }\right),
\label{eq:wind number density}
\end{eqnarray}
%
where $\dot{\Sigma}_{\rm sf}=\Sigma_{\rm H_2}/\tau_{\rm sf}$ is used. The
density is consistent with the observed diffuse thermal X-rays under the
assumed temperature of $k_{\rm B}T_{\rm w}\simeq0.3$~keV. Note that the
crossing time of the blown gas over the layer, $z_{\rm hl}/C_{\rm s,w}
\simeq10$~Myr, and $n_{\rm w}\simeq0.4\times10^{-3}~{\rm cm^{-3}}$ may give
ionization states of the thermal X-ray emission consistent with
the observation \citep{yamamoto22}.

\subsection{the Galactic cosmic rays}
\label{sec:cr}
The blown gas can be accelerated by the CR pressure and extend
to a region of $z>z_{\rm hl}$. In this paper, we mainly focus on relatively
low-energy ($\sim$~GeV) CRs, which dominate the CR energy density.
\par
The CR transport equation is simply written as
%
\begin{eqnarray}
\frac{\partial {\cal N}_{\rm cr}(\bm{r},\gamma) }{\partial t}
=\dot{{\cal N}}_{\rm cr,s}(\bm{r},\gamma)
+ {\cal D}_{\rm cr}(\gamma)\bm{\nabla}^2 {\cal N}_{\rm cr}(\bm{r},\gamma),
\label{eq:cr_trans}
\end{eqnarray}
%
where ${\cal D}_{\rm cr}(\gamma)$ is a diffusion coefficient of CRs and $\gamma$
is the Lorentz factor of CRs, respectively. The CR injection rate density
$\dot{N}_{\rm cr,s}$ is discussed later.
For simplification, here we neglect the CR advection, momentum diffusion, and non-trivial interaction
with the background fluid. The fully self-consistent treatment for those effects has not been
well established yet,
though, several recent attempts have developed the numerical scheme
\citep[e.g.][]{girichidis20,girichidis22,girichidis24,2023MNRAS.521.3023T}.
As will be shown later, our treatment results in a soft CR spectrum.
However, maintaining the total energy budget regulated by the star
formation activities, we mainly focus on the 1~GeV gamma-ray brightness below. 
The caveats for the adopted simplification in our method will be discussed in
Section~\ref{sec:discussion}. We also discuss the effects of advection in
the appendix~\ref{app}.
\par
With the steady-state approximation, the
formal solution for the CR density ${\cal N}_{\rm cr}$ becomes
%
\begin{eqnarray}
{\cal N}_{\rm cr}(\bm{r},\gamma)
&\approx&
\frac{\tau_{\rm cr}(\gamma)}{4\pi H^2}
\int
\frac{ \dot{{\cal N}}_{\rm cr,s}(\bm{r}',\gamma) }{ |\bm{r}-\bm{r}'| }
d^3\bm{r}',
\label{eq:cr_trans_formal}
\end{eqnarray}
%
where $\tau_{\rm cr}=H^2/{\cal D}_{\rm cr}(\gamma)$ is the residence
time of CR at the disk (the appendix~\ref{app}).
The CR composition of unstable radioactive
isotopes such as Be$^{10}$ implies that the residence time of the
low-energy CRs is $\sim1$~Myr~\citep[e.g.,][for recent reviews]{gabici19}.
In this paper, we assume the residence time as
%
\begin{eqnarray}
\tau_{\rm cr}=1~{\rm Myr}\left(\frac{\gamma}{2}\right)^{-0.6},
\end{eqnarray}
%
for simplicity, which is equivalently ${\cal D}_{\rm cr}\simeq
2.7\times10^{28}~{\rm cm^2~s^{-1}}(\gamma/2)^{0.6}(H/0.3~{\rm kpc})^2$.
The CR injection rate density is defined as
%
\begin{eqnarray}
\dot{{\cal N}}_{\rm cr,s}
=\dot{\cal N}_0\left( \frac{ \gamma }{2} \right)^{-2.1}
~~~(\gamma\ge2),
\end{eqnarray}
%
where the normalization factor is given by the energy injection rate
density as $\dot{e}_{\rm cr,s}=\int_2^{\infty}\dot{{\cal N}}_{\rm cr,s}
\gamma m_{\rm p}c^2d\gamma$, where $c$ is the speed of light.
In equation (\ref{eq:cr_trans}), we neglect
the energy loss due to $pp$-collision, whose timescale can be
estimated as $1/(K_{pp} n\sigma_{\rm h} c)\sim
100~{\rm Myr}(n/1~{\rm cm^{-3}})^{-1}$
\citep[e.g.,][]{schlickeiser02}, where the inelasticity
$K_{pp}\sim 0.3$ and $\sigma_{\rm h}\sim 3 \times 10^{-26}$~cm$^2$ is used.
The energy injection rate density is obtained from the supernova rate as
%
\begin{eqnarray}
\dot{e}_{\rm cr,s}=
\frac{ \eta_{\rm cr}E_{\rm sn} }{ \bar{m}_{\rm *,ms} }
\dot{\Sigma}_{\rm sn}(R)\delta(z),
\end{eqnarray}
%
where the CR injection efficiency is $\eta_{\rm cr}=0.1$, and $\delta(z)$ is
the Dirac's delta function. The CR energy density and pressure are given
by $e_{\rm cr}(R,z)=\int_2^\infty{\cal N}_{\rm cr}\gamma m_{\rm p}c^2d\gamma$
and $P_{\rm cr}=e_{\rm cr}/3$, respectively.
\par
\begin{figure}[htbp]
\includegraphics[scale=1.1]{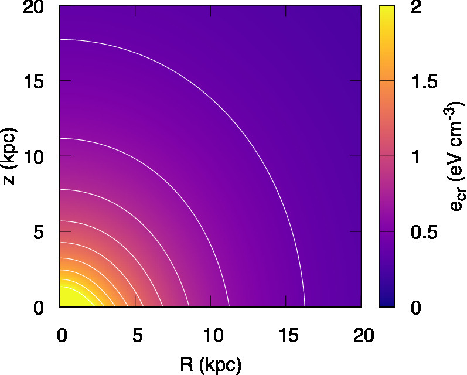}
\caption{The CR energy density in the Galactic halo. The contours show
the density increasing by 0.2~eV~cm$^{-3}$ from $0.4$~eV~cm$^{-3}$
to 2.0~eV~cm$^{-3}$.}
\label{fig:ecr}
\end{figure}

Our steady solution results in
${\cal N}_{\rm cr}\propto\gamma^{-2.7}$ everywhere and the spectral index
of $-2.7$ is consistent with the observed CR energy distribution around the Earth \citep[e.g. one of the latest results:][]{2022PhRvL.129j1102A}.
Figure~\ref{fig:ecr} shows the calculated CR energy density in the Galactic halo.
The energy density at $R\simeq8.5$~kpc is $\simeq0.8$~eV~cm$^{-3}$, which agrees
with the observed energy density around the Earth.
%
%

\subsection{the Galactic halo}
\label{sec:wind}
Here we describe our model for the Galactic wind dynamics. The observed halo
gas consists of hot, tenuous X-ray emitting gas \citep[e.g.,][]{predehl20}
and cold, dense HI clouds \citep[e.g.,][]{wakker97}. The gas dynamics should
obey highly nonlinear processes as seen in the interstellar medium
\citep[ISM, e.g.,][]{koyama02}. To describe their dynamics precisely, we should
take into account the nonlinear processes of magnetohydrodynamics by including
the effects of CRs, ionization structure, thermal conduction, radiative cooling,
heating process, and so on. While the full numerical simulations of the halo winds
require a high computational cost \citep[e.g.,][]{brent23,brent24,armillotta24},
the global structure of the halo gas we are interested in mainly depends on
the pressure balance among gravity, CR pressure, and magnetic pressure
\citep[e.g.,][]{boulares90,ferriere01}.
\par
\citet{shimoda22a} studied the effects of CRs for launching the Galactic wind
and found that only when the temperature of the wind gas is kept high by the
CR heating, the wind is successfully launched. In the Galactic halo, the gas
heating rate due to the dissipation
of Alfv\'en waves induced by CRs,
$Q_{\rm w}=V_{\rm A}|\bm{\nabla}P_{\rm cr}|$, can be comparable to the radiative
cooling rate of $n_{\rm w}{}^2\Lambda$;
%
\begin{eqnarray}
\frac{n_{\rm w}{}^2\Lambda}{Q_{\rm w}}
&&
\simeq0.9
\left( \frac{ n_{\rm w}    }{ 10^{-3}~{\rm cm^{-3}}          }\right)^{5/2}
\left( \frac{ \Lambda      }{ 10^{-22}~{\rm erg~cm^3~s^{-1}} }\right)
\nonumber \\
&&
\left( \frac{ B            }{  1{\rm \mu G}                 }\right)^{-1}
\left( \frac{ e_{\rm cr}   }{  1~{\rm eV~cm^{-3}}           }\right)^{-1}
\left( \frac{ H_{\rm cr}   }{ 10~{\rm kpc}                  }\right),
\label{eq:ratio-clht}
\end{eqnarray}
%
where $V_{\rm A}$ is the Alfv\'en velocity, and $B$ is the magnetic field.
When the CR heating rate is larger than the radiative cooling rate, the gas
temperature maintains at the virial temperature, going to a larger $z$, and
finally escaping from the Galactic system as the wind by the CR pressure.
\par
To save the computational costs, we omit solving these cooling and heating
processes, and simplify the halo gas dynamics as follows.\footnote{The magnetic
field strength is one of the most uncertain quantities in the Galactic halo.
Even if we suppose a comparable strength to the strength at the disk as
$B\sim1$~${\rm \mu}$G, the plasma beta becomes
$\beta\sim10$
$(n_{\rm w}/10^{-3}~{\rm cm^{-3})}$
$(kT_{\rm w}/0.3~{\rm keV})$
$(B/1~{\rm \mu G})^{-2}$
and the magnetic pressure becomes
$\sim0.02~{\rm eV~cm^{-3}}$$(B/1~{\rm \mu G})^{2}$
$< P_{\rm cr}$.}
The validity of this simplification will be justified later.
The equation of motion of the fluid element can be approximately written as
%
\begin{eqnarray}
\frac{d\bm{v}_{\rm w}}{dt}=-\frac{1}{\rho_{\rm w}}\bm{\nabla}P_{\rm cr}-\bm{g},
\label{eq:wind}
\end{eqnarray}
%
where $\bm{v}_{\rm w}$ and $\bm{g}$ are the velocity and gravitational
acceleration vectors, respectively. From the force balance along the $z$
direction,$-\partial P_{\rm cr}/\partial z-\rho_{\rm w}g_z>0$, the criterion
density for the escaping outflow can be written as
%
\begin{eqnarray}
n_{{\rm w,G}}
&&< 0.4\times10^{-3}~{\rm cm^{-3}}
\nonumber \\
&&
\left( \frac{ e_{\rm cr} }{  1~{\rm eV~cm^{-3}}               } \right)
\left( \frac{ H_{\rm cr} }{ 10~{\rm kpc}                      } \right)^{-1}
\left( \frac{ \Sigma_*   }{ 250~M_\odot~{\rm pc^{-2}}         } \right)^{-1},
\end{eqnarray}
%
where we have adopted $\partial P_{\rm cr}/\partial z\sim
- P_{\rm cr}/H_{\rm cr}$, and $g_z\sim2\pi\Sigma_*$. The threshold mass
density of the stellar disk is estimated as $\Sigma_*\sim M_*/(2\pi
R_{*,{\rm d}}^2)\sim 250~M_\odot~{\rm pc^{-2}}$, where $M_*\sim4\times
10^{10}~M_\odot$ and $R_{*,{\rm d}}\sim5$~kpc \citep[e.g.,][]{blandh16}.
An outflow from the inner region ($R\lesssim5$~kpc) where the number density
is larger than the above value, can not escape from the Galactic system, falling
back onto the Galactic disk, similar to the Galactic fountain flow.
This critical density is close to the critical density for the cooling effect
shown in eq. (\ref{eq:ratio-clht}). Therefore, our method with eq. (\ref{eq:wind})
can roughly reproduce the escaping wind and fallback gas without considering
the cooling effect. Considering the strong density dependence of $n_{\rm w}{}^2
\Lambda/Q_{\rm w}\propto n_{\rm w}{}^{5/2}$, we can reasonably conclude that
the gas dynamics mainly depends on $n_{\rm w}$ and the CR pressure.\footnote{When
$P_{\rm cr}$ is small, the thermal pressure becomes negligible due to the radiative
cooling. The cooling time scale is
$\tau_{\rm cool}\sim k_{\rm B}T_{\rm w}/(n_{\rm w}\Lambda)
\sim160$~Myr
$(k_{\rm B}T_{\rm w}/0.3~{\rm keV})$
$(n_{\rm w}/10^{-3}~{\rm cm^{-3}})^{-1}$
$(\Lambda/10^{-22}~{\rm erg~cm^3~s^{-1}})^{-1}$, while the dynamical time scale is
$\tau_{\rm dyn}\sim H_{\rm cr}/v_{\rm w}\sim 50$~Myr$\left(H_{\rm cr}/10~{\rm kpc}\right)
\left(v_{\rm w}/200~{\rm km~s^{-1}}\right)^{-1}$.}
An outflow from $R\gtrsim 5$~kpc can escape from the Galactic system as
the wind, but it is relatively less important for the thermal X-ray emissivity
and hadronic gamma-ray emissivity.
\par
The Galactic fountain-like flow
implies the coexistence of the outflow and infalling failed wind in the halo.
From the condition of $n_{\rm w,c}\sim n_{\rm w,G}$, we expect that the failed
wind suffers the radiative cooling and becomes condensed gas due to the thermal
instability \citep[e.g.,][]{field65,shapiro76}. Such multiphase gas, in which
cold, dense gas coexists with hot, diffuse gas, is ubiquitously observed in
the ISM, in the MW halo \citep[e.g.,][]{bregman07,das19}, and in the external
galaxy's halo \citep[e.g.,][for reviews]{tumlinson17}. Moreover, many HI clouds
falling into the MW disk are observed
\citep[the so-called high-velocity/intermediate-velocity cloud, e.g.,][]{wakker97}.
In this paper, we assume that the halo gas density is highly structured in reality
and the motion of the outflowing wind is not affected by
the inflowing gas (like "raindrops" in the atmosphere).
\par
We summarize our expectations as follows. The outflow from the inner
region ($R\lesssim5$~kpc and $n_{\rm w}\gtrsim10^{-3}$ cm$^{-3}$)
suffers radiative cooling and falls back onto the disk. The motion of
the fluid element can be approximated by the equation~\eqref{eq:wind}.
The outflow from the outer region ($R\gtrsim5$~kpc and $n_{\rm w}\lesssim
10^{-3}$ cm$^{-3}$) is affected by the CR heating so that the thermal pressure
is large enough. However, a lower density in the outer wind may result in lower
emissivities of the thermal X-ray and hadronic gamma-ray. Even if we adopt the
equation~\eqref{eq:wind} to their dynamics, the X-ray and gamma-ray intensities
are less affected. The outflow at the critical density
$n_{\rm w}\simeq10^{-3}$ cm$^{-3}$ ($R\simeq5$~kpc) can be regarded as
a minor component because of $n_{\rm w,c}\propto n_{\rm w}^{5/2}$ and
the exponential profile of $\Sigma_{\rm H_2}$
(see, equations~\ref{eq:H2} and \ref{eq:wind number density}).
\par
For numerical calculation of the wind, following \citet{shimoda24}, we treat
the fluid element as a test particle denoted by subscript $i$ and adopt the
equation~(\ref{eq:wind}) as its equation of motion under the axisymmetric
approximation. We prepare the particles at $(R_i,z_{\rm hl})$, where $R_i$ is
discretized with the interval of $\Delta R=(30/256)~{\rm kpc}\simeq0.017~{\rm kpc}$).
Considering the criterion for the fallback discussed above, we can simplify
the estimate of the acceleration by the CR pressure using the initial density
$\rho_{\rm w,hl}(R_i,z_{\rm hl})$. We set two spatial boundary conditions at
$r=30$~kpc as the free escape boundary and $z=0.3$~kpc, below which the wind
merges with the disk gas. The initial velocity components are set to be
$(v_{{\rm w},R},v_{{\rm w},\phi},v_{{\rm w},z})
=(0,v_{{\rm w},\phi,i}(R_i,z_{\rm hl}), C_{\rm s,w})$, where the initial rotation
velocity $v_{\rm w,\phi,i}$ is a free parameter in this model. We use simple
formulae of the gravitational potentials due to the stars \citep{miyamoto75} and DMs \citep{nfw96} following \citet{shimoda22a}.
\par
We have adjusted the model parameters to reproduce the current MW conditions except
for the systematic rotation of the halo gas, which is currently not constrained.
The angular momentum transportation is an important factor for the long-term
evolution of galaxies in general.
On the other hand, the angular momentum (AM) can be
redistributed at the disk-halo interface by
e.g., magnetic fields \citep{breitschwerdt99,kakiuchi24},
and a large-ordered field possibly regulates the outflow
\citep{meliani24}. In this paper,
we test two cases parametrizing the initial rotation velocity of the wind as
$v_{{\rm w},\phi,i}(R_i,z_{\rm hl})=\sqrt{R_ig_{R}(R_i,z_{\rm hl})}$ (co-rotating halo)
and $v_{{\rm w},\phi,i}(R_i,z_{\rm hl})=0.5\sqrt{R_ig_{R}(R_i,z_{\rm hl})}$
(AM redistribution case).
\par
%
\begin{figure}[htbp]
\includegraphics[scale=1.0]{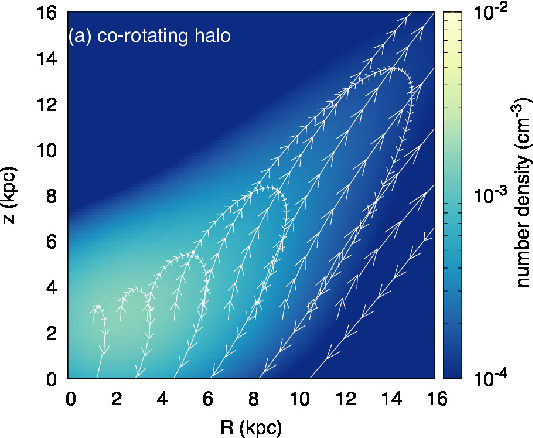}
\includegraphics[scale=1.0]{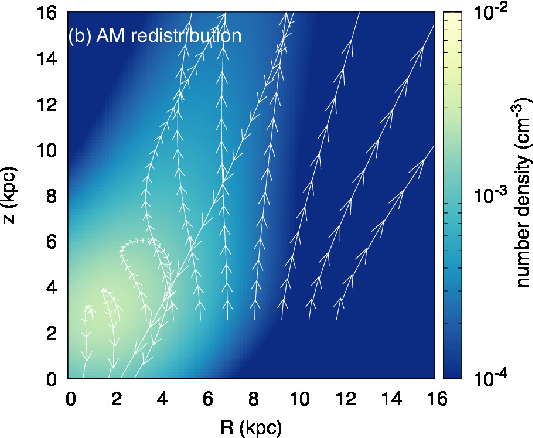}
\caption{The trajectories of the fluid elements (vectors). The color map
shows the wind density. The panels (a) and (b) show the case of
$v_{{\rm w},\phi,i}(R_i,z_{\rm hl})
=\sqrt{R_i g_R(R_i,z_{\rm hl})}$ (co-rotating halo) and
$v_{{\rm w},\phi,i}(R_i,z_{\rm hl})
=0.5\sqrt{R_i g_R(R_i,z_{\rm hl})}$ (AM redistribution), respectively.}
\label{fig:wind}
\end{figure}
%
Figure~\ref{fig:wind} shows the trajectories of the wind particles. Assuming
continuous wind ejections and steady winds, we obtain the wind density from
the particle trajectories with the mass conservation and Gaussian smoothing
with the width of $z_{\rm hl}$. In the co-rotating halo case (Figure~\ref{fig:wind}a),
the centrifugal force equilibrium condition, $v_{{\rm w},\phi,i}=
\sqrt{R_ig_R(R_i,z_{\rm hl})}$, results in a simple fountain flow; the wind goes
in the outward direction due to the centrifugal force and eventually falls back
on the disk. In the AM redistribution case shown in
Figure~\ref{fig:wind}b,
$v_{{\rm w},\phi,i}=0.5\sqrt{R_ig_R(R_i,z_{\rm hl})}$, the wind is concentrated
toward the inner Galactic region attracted by the stellar gravity. Although our
treatment neglects the effects depending on $B$ and $\Lambda$, the numerical results
are consistent with our expectations discussed above. 

\subsection{the thermal X-rays and hadronic gamma-rays}
For the thermal X-ray emission from the halo in the {\it eROSITA} $0.6$-$1$~keV
band, we assume an average, spatially uniform emissivity of $\Lambda_{\rm X}
=10^{-23}~{\rm erg~cm^3~s^{-1}}$ for simplicity. This assumption is equivalent
to the isothermal gas with a temperature of $\sim 0.3$ keV. The hadronic gamma-ray
emission from CRs via $pp$-collision is calculated with the same method in
\citet{Nishiwaki-paperI}. For the total cross-section, we use the formulae in
\citet{Kamae2006,Kamae2007}. The energy distribution of $\pi^0$ produced in a
collision is expressed by the formulae in \citet{Kelner2006}. With the
$\pi^0$-injection rate density $\dot{q}_\pi(E_\pi)$ obtained with the method above,
the gamma-ray emissivity due to $\pi^0$-decay is calculated as
%
\begin{eqnarray}
\dot{\varepsilon}(E_\gamma)
=2E_\gamma \int_{E_{\rm th}}
  \frac{\dot{q}_\pi(E_\pi)}{\sqrt{E_\pi^2-m_\pi^2 c^4}}dE_\pi,
\end{eqnarray}
%
where $E_{\rm th}=E_\gamma+m_\pi^2 c^4/(4E_\gamma)$. The intensity map is
obtained by integrating the emissivity along the line of sight.
\par
The order of magnitude estimate of the hadronic gamma-ray intensity,
${\cal J}_\gamma$, may be given by \citep{schlickeiser02}
%
\begin{eqnarray}
{\cal J}_\gamma
&\sim&
\frac{D}{4\pi}\frac{n_{\rm cr}n_{\rm w}\sigma_{\rm h}c}{m_{\pi}c^2}
\left[\frac{E_\gamma}{m_\pi c^2}+\frac{m_\pi c^2}{4E_\gamma}\right]^{-1.6}
\nonumber \\
&\sim&
0.7\times10^{-6}~{\rm ph~cm^{-2}~s^{-1}~GeV^{-1}~str^{-1}}
\nonumber \\
&&
\left(  \frac{  D          }{   3~{\rm kpc}         } \right)
\left(  \frac{  e_{\rm cr} }{   1~{\rm eV~cm^{-3}}  } \right)
\nonumber \\
&&
\left(  \frac{  n_{\rm w } }{ 10^{-2}~{\rm cm^{-3}} } \right)
\left(  \frac{  E_\gamma   }{   1~{\rm GeV}         } \right)^{-1.6},
\label{eq:J estimate}
\end{eqnarray}
%
where $m_\pi=140$~MeV is the pion mass, $\sigma_{\rm h}\sim10^{-26}~{\rm cm^2}$
is the cross-section, $D$ is the path length, and we use
$n_{\rm cr}\sim e_{\rm cr}/m_{\rm p}c^2$. Note that ${\cal J}_\gamma
\propto\dot{\Sigma}_{\rm sf}{}^2$ in our model and the neutrino emission
with a comparable intensity to ${\cal J}_\gamma$ is also expected.

\section{Results}
\label{sec:results}
The numerical results of the thermal X-ray and hadronic gamma-ray
intensities are presented here. We compare
the surface brightness profiles to the actual observations and predict
the line-of-sight velocity distribution of each co-rotating halo case
and AM redistribution case for future radio and
X-ray observations.
\par
\begin{figure}
\includegraphics[scale=1.0]{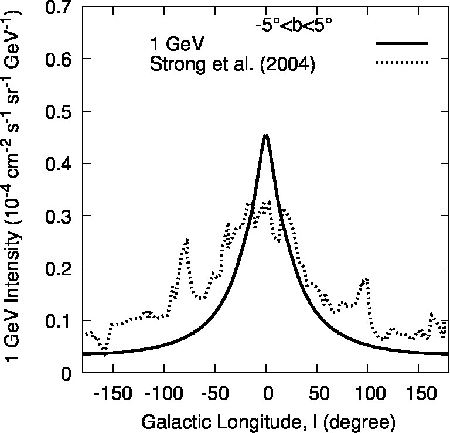}
\caption{The surface brightness of the 1~GeV hadronic gamma-ray
at the disk (solid line).
The dots are derived from \citet{strong04} as a guide.}
\label{fig:pp_disk}
\end{figure}
%
Figure~\ref{fig:pp_disk} shows the surface brightness of the 1~GeV hadronic gamma-ray
from the disk ($|b|<5^\circ$). This shows that the CR density we assumed consistently reproduces the observed gamma-ray brightness at
the disk \citep[e.g.,][]{strong04}.
\par
%
\begin{figure}[htbp]
\includegraphics[scale=0.8]{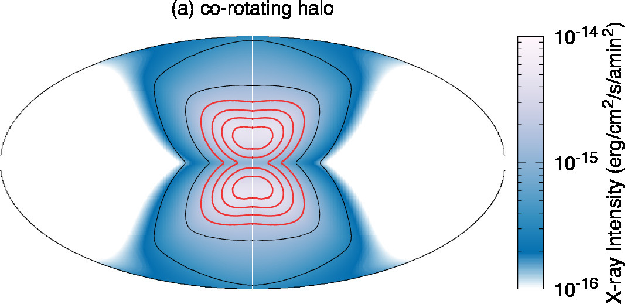}
\includegraphics[scale=0.8]{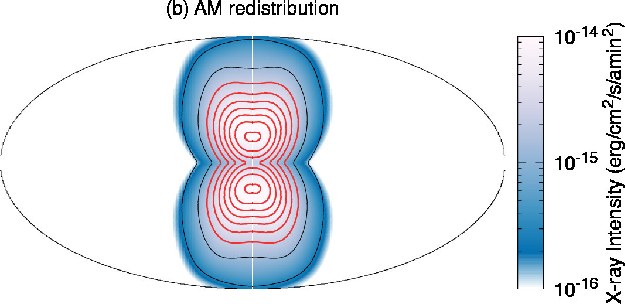}
\caption{The intensity sky maps of the hadronic gamma-ray and X-ray observed
from $(R,z)=(8.5~{\rm kpc},0~{\rm kpc})$. The contour shows the 1~GeV gamma-ray
intensity (${\rm photon~cm^{-2}~s^{-1}~str^{-1}}$) increasing
by $0.25\times10^{-6}$ from $0$ to $3\times10^{-6}$. The red contours indicate
the intensity larger than $0.5\times10^{-6}$.
The color shows the X-ray intensity for the $0.6$-$1$~keV band under
the assumed emissivity of $n_{\rm w}\Lambda_{\rm X}/4\pi$,
where $\Lambda_{\rm X}=10^{-23}~{\rm erg~~cm^3~s^{-1}}$.}
\label{fig:sky}
\end{figure}
%
As shown in Figure~\ref{fig:wind}, the failed wind around the inner
region may be observed as `bubbles' in the X/$\gamma$-ray sky due to
the projection effect. Figure~\ref{fig:sky} shows the estimated intensity maps of
the thermal X-ray and hadronic gamma-ray at $E_\gamma=1$~GeV.
Here we only calculate the emission from the halo.
The numerically computed ${\cal J}_\gamma$ at $E_{\gamma}=$1~GeV
is consistent with the estimated intensity given by the equation~\eqref{eq:J estimate}
and the observed intensities \citep[e.g.,][]{su10,ackermann14}.
The morphological relation between the eRBs and FBs is that
the FBs are bright {\it inside} the eRBs. Such morphology is
reproduced very well in both the two models for the initial rotation.
\par
%
%
\begin{figure}[htbp]
\includegraphics[scale=1.0]{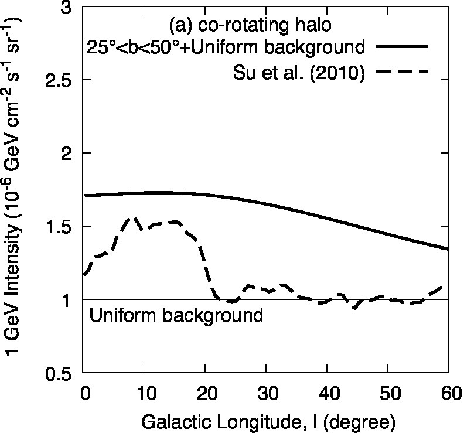}
\includegraphics[scale=1.0]{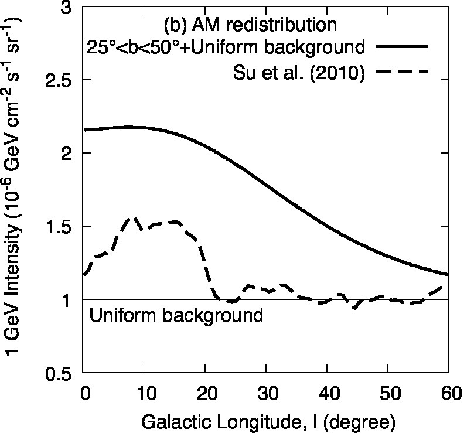}
\caption{
The surface brightness profile of the 1~GeV hadronic gamma-ray observed
from $(R,z)=(8.5~{\rm kpc},0~{\rm kpc})$. We add the uniform background component
with brightness of $10^{-6}~{\rm GeV~cm^{-2}~s^{-1}~sr^{-1}}$. The broken line
is derived from \citet{su10} as a guide.}
\label{fig:pp_profile}
\end{figure}
%
Figure~\ref{fig:pp_profile} shows the surface brightness profile of the hadronic
gamma-ray averaged on the Galactic latitude ranges of $25^\circ<b<50^\circ$.
The actual profiles of the FBs are analyzed by \citet{su10} (Figure 8) and
\citet{ackermann14} (Figure 23) for examples. The surface brightness of the FBs
is not perfectly symmetric; the northern bubble is $\sim2$ times brighter than
the southern bubble. Here we compare our results to the southern bubble where
the gamma-ray foreground is fainter than the northern sky \citep{su10}.
Our model results in consistent
surface brightness with the observations.
Note that our primitive model does not aim to exactly reproduce the gamma-ray flux.
The difference with a factor $\sim2$ does not matter in this paper.
The surface brightness profile flattens at $|l|\lesssim20^\circ$, which is one of
the non-trivial features of the FBs.
The AM redistribution case results in such
a flat profile, while the co-rotating
halo case results in a shallower profile. Thus,
the AM redistribution case is favored
for explaining the FBs.
Compared to the observed sharp drop of the gamma-ray intensity at the edge of
the FBs at $\l=20$ degree, the model intensity gradually decreases with the Galactic
longitude for $\l>20$ degree.
\par
%
\begin{figure}[htbp]
\includegraphics[scale=1.0]{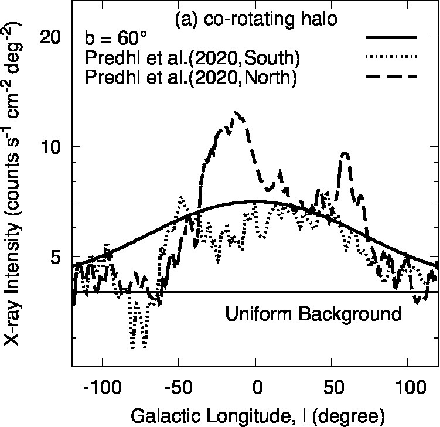}
\includegraphics[scale=1.0]{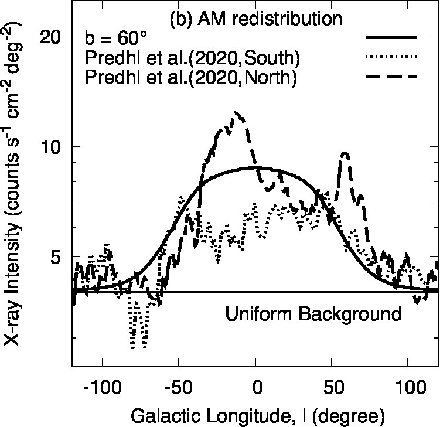}
\caption{
The surface brightness profile of the X-ray at $b=60^\circ$ observed
from $(R,z)=(8.5~{\rm kpc},0~{\rm kpc})$ (the solid line).
The dots and the broken line are referred from
\citet{predehl20} as guides for the southern and northern bubbles, respectively.
The intensity is scaled as $1$~counts~s$^{-1}$~cm$^{-2}$~deg$^{-2}
=2.7\times10^{-16}$~erg~s$^{-1}$~cm$^{-2}$~arcmin$^{-2}$ for {\it eROSITA}
$0.6$-$1$~keV band. We assume a uniform background with an intensity of $4$~counts~s$^{-1}$~cm$^{-2}$~deg$^{-2}$.}
\label{fig:X-ray_profile}
\end{figure}
%
Figure~\ref{fig:X-ray_profile} shows the surface brightness profiles of
the thermal X-ray at the Galactic latitudes of
$b=60^\circ$~\citep[see,][for a comparison]{predehl20}.
Here we assume a uniform background with an intensity of
$4$~counts~s$^{-1}$~cm$^{-2}$~deg$^{-2}$.
Similar to the FBs, the observed surface brightness of the eRBs
is not perfectly symmetric; the northern bubble is $\sim2$ times
brighter than the southern bubble.
The surface brightness of the observed northern and southern bubbles is
$\gtrsim6$~counts~s$^{-1}$~cm$^{-2}$~deg$^{-2}$ at $-45^\circ\lesssim l\lesssim
65^\circ$ and shows flat profiles respecting $l$. The co-rotating halo case
shows smaller $l$ gradients than the observed northern bubble.
The AM redistribution case may be preferred
for explaining the northern bubble, while both cases
are consistent with the southern bubble.
\par
%
\begin{figure}[htbp]
\begin{center}
\includegraphics[scale=0.7]{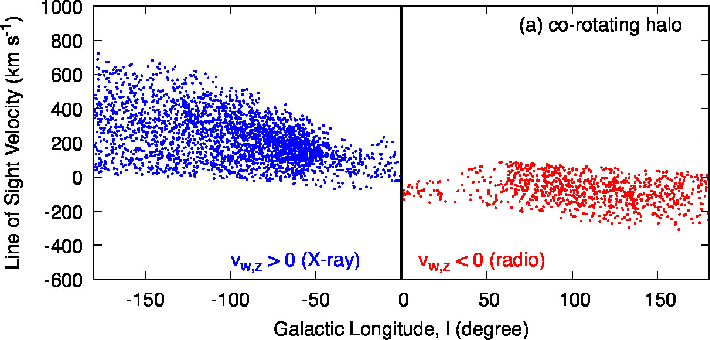}
\includegraphics[scale=0.7]{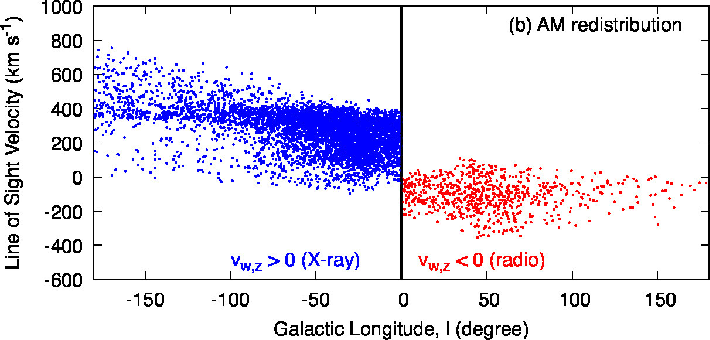}
\end{center}
\caption{The line of sight velocities of the halo gas at $40^\circ<b<80^\circ$
observed from $(R,z)=(8.5~{\rm kpc},0~{\rm kpc})$ under the assumed
rotation velocity of the solar cycle of $230~{\rm km~s^{-1}}$.
The case $v_{{\rm w},z}<0$ is shown at $l>0$ and case $v_{{\rm w},z}>0$
is shown at $l<0$.}
\label{fig:los}
\end{figure}
%
%
\begin{figure}[htbp]
\begin{center}
\includegraphics[scale=0.7]{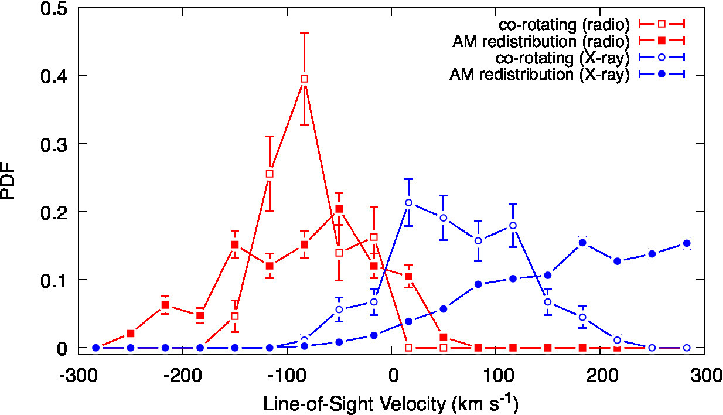}
\end{center}
\caption{The probability distribution functions of the
line of sight velocities of the halo gas at $40^\circ<b<80^\circ$
and $|l|<50^\circ$. The 1$\sigma$ statistical errors due to
a finite number of the samples are also shown. The unfilled (filled)
circles/squares indicate the co-rotating (AM redistribution) case.}
\label{fig:pdf}
\end{figure}
%
Figure~\ref{fig:los} shows the line-of-sight velocity distribution of
the fluid elements for $40^\circ<b<80^\circ$. The azimuthal position of
each wind particle is randomly selected. Here we regard that the outflow
particles ($v_{z,{\rm w}>0}$) are observed at the X-ray band and that the
inflow particles ($v_{z,{\rm w}}<0$) are cooled. The inflowing cold gas
is assumed to be bright at the radio band (the $21$~cm line emission) and
observed like the high velocity/intermediate velocity HI clouds
\citep[e.g.,][]{wakker97,ashley22,hayakawa22}.
The sample density respecting $l$ reflects the gas angular momentum trivially;
the AM redistribution case results in centrally concentrated
samples. For a small longitude $l\lesssim50^\circ$, the line-of-sight velocity at
the radio band is bounded at $|v_{\rm los}|\lesssim 100~{\rm km~s^{-1}}$ in
the co-rotating halo case ($|v_{\rm los}|\lesssim 200~{\rm km~s^{-1}}$ in
the AM redistribution case).
The probability distribution functions of the line-of-sight velocity
at $40^\circ<b<80^\circ$ and $|l|<50^\circ$ are shown in Figure~\ref{fig:pdf}.
The radio observations of the high-velocity clouds around the FBs reported
the line-of-sight velocity range of $\sim\pm170~{\rm km~s^{-1}}$~\citep[e.g.,][]
{ashley22}, which is consistent with the AM redistribution case. Thus, the halo gas
dynamics will be tested by spectroscopy at the radio band and X-ray band.
In particular, we need a high-energy-resolution X-ray spectroscopy like
{\it XRISM} mission \citep{xrism20} and {\it Athena} mission \citep{barret18}.

\section{Conclusions and Discussion}
\label{sec:discussion}
We have shown that the eRBs and FBs can be explained by the persistent Galactic
wind scenario, which is well motivated by the long-term evolution of the MW;
the gas supply and consumption due to the star formation and Galactic wind are
balanced, keeping the total metal content in the disk \citep{shimoda24}.
As the computational cost for the halo dynamics simulations is high, we have modeled
the gas density profile affected by the disk wind with a very simplified method using
test particles. Despite this primitive estimate for the halo density profile, the
brightness and the spatial scales of the FBs and eRBs are roughly reproduced. Especially,
the non-trivial configuration, the FBs surrounded by the larger eRBs, is consistently
reproduced. In addition, the velocity distribution of the inflowing gases in our
model is similar to the observed velocities for high-velocity clouds. The results
imply that the steady Galactic fountain flow consistent with the star formation
history in the Galactic disk produces the dense halo structure, which is responsible
for both the FBs and eRBs. In this model, the FBs and eRBs are quasi-steady structures
due to the quasi-steady star formation activity in the Galactic disk.
The results prefer the case that the angular momentum of the wind is
partially extracted by some kind of dissipative process.
\par
On the other hand, our model does not reproduce the sharp edge of the FBs
and their hard spectra. Our steady solution for the CR transport
equation~(\ref{eq:cr_trans}) leads to
a soft gamma-ray spectrum as ${\cal J}_\gamma\propto E_{\gamma}{}^{-1.6}$ everywhere,
although the spectra of the FBs are as hard as $\propto E_{\gamma}{}^{-1}$.
The coexistence of the outflowing hot gas and infalling cold gas in the halo
implies hydrodynamical instabilities. A detailed modeling with the effects of
such instabilities may improve the results.
If we consider the re-acceleration mechanism
such as the stochastic acceleration by turbulence, the hard spectrum can
be reproduced. In our AM redistribution case, the halo
gas is concentrated towards the inner halo region. Thus, the inner region
can be selectively disturbed, and we naturally expect an efficient re-acceleration.
\par
The stochastic acceleration can also re-accelerate CR electrons \citep[see e.g.][]{1989ApJ...336..243S,2011MNRAS.412..817B,2019ApJ...877...71T}.
CR electrons in the Galactic halo should be responsible for the bright GRLs.
If the re-acceleration of electrons works very well, the leptonic gamma-ray emission
can be comparably bright with the FBs \citep[e.g.,][]{sasaki15}. The efficiency of
the re-acceleration can be tested by observing a hard X-ray synchrotron emission, although
there is only an upper limit on the hard X-ray intensity for the Galactic halo
\citep[e.g.,][]{kataoka13,kataoka18}. Thus, further studies for the re-acceleration and
observations at the hard X-ray band like {\it FORCE} mission \citep{force22}
are ideal. It should be emphasized that observations of the FBs
at the very high energy gamma-ray band by, e.g., the {\it Cherenkov Telescope Array
Observatory} and of neutrinos by, e.g., the {\it IceCube} experiment,
are also important to reveal the maximum energy of CRs
and to verify the hadronic/leptonic scenarios \citep[e.g., acceleration efficiency
ratio of CR protons to CR electrons, see,][]{Nishiwaki-paperI}.
\par
\citet{carretti13} estimated the magnetic energy of the GRLs with the
leptonic scenario for the gamma-ray intensity
as $U_{\rm B,lobe}\sim10^{55}$~erg. This is comparable to the energy
in the disk within the active star formation region; $U_{\rm B,disk}\sim
(B^2/8\pi)\times 2\pi R^2 H
\sim10^{55}~{\rm erg}(B/1~{\rm \mu G})^2(R/5~{\rm kpc})^2(H/300~{\rm pc})$.
In our scenario, the plasma circulates through the disk and the halo. Therefore, the
comparable magnetic energy may be acceptable. Note that even if the magnetic
reconnection  dissipates the magnetic energy,
which is one of the possible origins of the AM redistribution,
the halo gas concentration can be responsible for the turbulent dynamo at the halo.
This should be studied with stochastic acceleration.
\par
The morphology of the eRBs and FBs depends on the angular momentum of the halo gas in our
model. These are new important clues to study the long-term evolution
of the MW on the time scale of $\sim1$~Gyr.
Since the angular momentum of the disk gas is recorded to the formed stars, the
evolution of the angular momentum can be studied through the stellar dynamics. Thus,
there is a possible synergy between high-energy astrophysics, CR physics, and the
Galactic archaeological study of stellar dynamics. We will extend our model along
the lines of this study.

\acknowledgments
The authors thank S.-i. Inutsuka and M. Nagashima for fruitful discussions. We also thank the anonymous referee, for his/her comments that further improved the paper. This work is supported by the joint research program of the Institute for Cosmic Ray Research (ICRR), the University of Tokyo, and KAKENHI grant Nos. 22K03684, 23H04899, 24H00025 (K.A.), and 24K00677 (J.S.).


\appendix
\section{The advection effects on the CR transportation}
\label{app}
We evaluate the effects of advection, which is neglected in the CR transport
equation~\eqref{eq:cr_trans}. As discussed Sections \ref{sec:wind}
and \ref{sec:discussion}, the expected velocity field consists of inflows and outflows,
having a highly complicated structure in reality.
Such a velocity field cannot be treated fully consistently by our
two-dimensional model, in which all values are averaged out in the azimuthal angle
direction. To check the validity or limit of our treatment, we consider an
extreme case with a constant velocity field along the vertical direction,
$\bm{v}=(0,0,v_0)$ ($z>0$), which maximizes the advection effect.
Then, the transport equation becomes
%
\begin{eqnarray}
\frac{\partial {\cal N}_{\rm cr}}{\partial t}
=\dot{{\cal N}}_{\rm cr,s}
+ {\cal D}_{\rm cr}(\gamma)\bm{\nabla}^2 {\cal N}_{\rm cr}
-v_0\frac{\partial {\cal N}_{\rm cr}}{\partial z},
\label{eq:cr_trans_ad}
\end{eqnarray}
%
and its formal solution can be derived using the Green function as
%
\begin{eqnarray}
{\cal N}_{\rm cr}(t,\bm{r},\gamma)
=
\frac{1}{8\left(\pi {\cal D}_{\rm cr}\right)^{3/2}}
\int d^3\bm{r'}\int_{t_0}^{t}dt'
\frac{\dot{{\cal N}}_{\rm cr,s}(\bm{r'})}{\left(t-t'\right)^{3/2}}
\exp\left[
-\frac{ (x-x')^2 + (y-y')^2 +\left\{ (z-z')_-v_0(t-t') \right\}^2 }
       {4{\cal D}_{\rm cr}(t-t')}
\right].
\end{eqnarray}
%
Introducing a variable
$w=\sqrt{{\cal T}_{\rm ad}/(t-t')}$, where ${\cal T}_{\rm ad}
=4{\cal D}_{\rm cr}/v_0{}^2$, we obtain
%
\begin{eqnarray}
{\cal N}_{\rm cr}(t,\bm{r},\gamma)
=
\frac{1}{4\left(\pi {\cal D}_{\rm cr}\right)^{3/2}}
\int d^3\bm{r'}\dot{{\cal N}}_{\rm cr,s}(\bm{r'})
\int_{w_0}^{\infty}
\frac{ dw }{ {\cal T}_{\rm ad}{}^{1/2}  }
\exp\left[
         -aw^2-\frac{1}{w^2}+\frac{2h}{v_0{\cal T}_{\rm ad}}
         \right],
\end{eqnarray}
%
where $h= z-z'$, $a=|\bm{r}-\bm{r'}|^2/(4{\cal D}_{\rm cr}{\cal T}_{\rm ad})$,
and $w_0=\sqrt{{\cal T}_{\rm ad}/(t-t_0)}$,
respectively. Since the advection time scale,
${\cal T}_{\rm ad}\sim10$~Myr$(\gamma/2)^{0.6}\left(v_0/C_{\rm s,w}\right)^{-2}$ 
with $C_{\rm s,w}=200$~km~s$^{-1}$, is significantly shorter than $(t-t_0)\ga 1$~Gyr,
we can approximate $w_0\approx0$. Then, the integration of $w$
is carried out by the Gaussian integral (or the Euler-Poisson integral) as
%
\begin{eqnarray}
{\cal N}_{\rm cr}(\bm{r},\gamma)
\approx
\frac{\tau_{\rm cr}(\gamma)}{4\pi H^2}
\int d^3\bm{r'}
\frac{ \dot{{\cal N}}_{\rm cr,s}(\bm{r'}) }{ |\bm{r}-\bm{r}'| }
\exp\left[ -\frac{ |\bm{r}-\bm{r'}| - 2(z-z') }{ {\cal H}_{\rm ad}(\gamma) } \right],
\label{eq:cr_trans_ad_formal}
 \end{eqnarray}
%
where ${\cal D}_{\rm cr}(\gamma)
=H^2/\tau_{\rm cr}(\gamma)$ is used (see the main text), and
${\cal H}_{\rm ad}(\gamma)
=4{\cal D}_{\rm cr}(\gamma)/v_0
\sim2$~kpc~$(\gamma/2)^{0.6}(v_0/C_{\rm s,w})^{-1}$ is
the scale height of the advection.
Compared with a pure diffusion case of the equation~\eqref{eq:cr_trans_formal},
the last exponential factor in the equation~\eqref{eq:cr_trans_ad_formal}
indicates the correction by the advection term.
When $|\bm{r}-\bm{r}'|\ll {\cal H}_{\rm ad}$, the factor can be reduced as unity,
and we obtain the equation~\eqref{eq:cr_trans_formal}.
This means that a contribution of a source at
$\bm{r}'=(x',y',0)$ for a CR number density
at a position of $\bm{r}=(x,y,z)$ is mostly determined by diffusion.
Our model focuses on $\sim1$~GeV gamma-rays emitted by CRs with
$\gamma\sim 20$. In this case, ${\cal H}_{\rm ad}\sim 8$~kpc is significantly
large even in this extreme/simplified setup for the velocity field. The transport
equation \eqref{eq:cr_trans} may not lead to a largely different CR density
around the GC within a radius of $\la8$~kpc, which is compatible with the size
of the FBs. As discussed in Section~\ref{sec:discussion}, the effects of a realistic,
turbulent velocity field will be studied in future work.

%






\bibliography{apj_sjsi22}{}
\bibliographystyle{aasjournal}

\end{document}